\documentclass[aps,prd,unsortedaddress,
  showpacs,preprintnumbers]{revtex4} 

\usepackage{graphicx}
\usepackage{dcolumn}
\newcommand{\be}{\begin{equation}}
\newcommand{\ee}{\end{equation}}
\newcommand{\bea}{\begin{eqnarray}}
\newcommand{\eea}{\end{eqnarray}}

\newcommand{\bfk}{\mbox{\boldmath $k$}}
\newcommand{\bfq}{\mbox{\boldmath $q$}}

\newcommand{\pup}{p^\uparrow}

\newcommand{\bfP}{\mbox{\boldmath $P$}} 
\newcommand{\bfS}{\mbox{\boldmath $S$}}

\newcommand{\kt}{k_\perp}
\newcommand{\bkt}{\mbox{\boldmath$k_\perp$}}
\def\lsim{\mathrel{\rlap{\lower4pt\hbox{\hskip1pt$\sim$}}\raise1pt\hbox{$<$}}}
\def\gsim{\mathrel{\rlap{\lower4pt\hbox{\hskip1pt$\sim$}}\raise1pt\hbox{$>$}}}

\begin{document}


\title{Sivers Effect and Transverse Single Spin Asymmetry in 
{\boldmath$e+p^\uparrow \to  e+J/\psi +X $\\}}

\author{Rohini M. Godbole}
\affiliation{Centre for High Energy Physics, Indian Institute of Science, Bangalore, India.}
\email{rohini@cts.iisc.ernet.in}

\author{Anuradha Misra}
\affiliation{Department of Physics, University of Mumbai, Mumbai, India.}
\email{misra@physics.mu.ac.in}

\author{Asmita Mukherjee}
\affiliation{Department of Physics, Indian Institute of Technology Bombay, Mumbai, India.} 
\email{asmita@phy.iitb.ac.in}

\author{Vaibhav S. Rawoot}
\affiliation{Department of Physics, University of Mumbai, Mumbai, India.}
\email{vaibhav.rawoot@physics.mu.ac.in}

\date{\today}
\begin{abstract}

We discuss the possibility of using electroproduction of $J/\psi $ as a probe of gluon Sivers function 
by measuring single spin asymmetry (SSA) in experiments with transversely polarized protons and electron beams. 
We estimate SSA for JLab, HERMES, COMPASS and eRHIC energies using the color evaporation model of 
charmonium production and find asymmetry up to 25 $\%$ for certain choices of model parameters which have been 
used earlier for estimating SSA in the  SIDIS and Drell-Yan processes. 

\end{abstract}

\pacs{13.88.+e, 13.60.-r, 14.40.Lb, 29.25.Pj}
\maketitle

\section{\label{sec:level1}Introduction}

In recent years, there has been a lot of interest in investigations of transverse 
single spin asymmetries (SSA's) in high energy QCD processes as they provide information 
about spin structure of proton. These asymmetries arise in  scattering of a transversely 
polarized proton off an unpolarized proton if the scattering cross section depends on 
the direction of polarization.

The single spin asymmetry (SSA) for the inclusive process $A^\uparrow + B \rightarrow C+X$ is defined as
\be 
A_N = \frac{d\sigma ^\uparrow \, - \, d\sigma ^\downarrow}
{d\sigma ^\uparrow \, + \, d\sigma ^\downarrow} \label{an}
\ee 
where $ d\sigma^{\uparrow(\downarrow)} $ denotes the cross section
for scattering of a transversely  polarized hadron A off an unpolarized hadron B,
with A upwards (downwards) transversely polarized with respect to the production plane.
Large SSA's have been measured in  pion production at Fermilab \cite{AdamsBravar1991} as well as 
at BNL-RHIC in $p p^\uparrow $ collisions~\cite{KruegerAllogower1999}. 
SSA's have also been observed by the HERMES \cite{Hermes} and COMPASS \cite{Compass} collaborations, 
in polarized semi-inclusive deep inelastic scattering. 
The magnitude of the observed asymmetries have been found to be larger than
what is predicted by pQCD \cite{alesio-review}.

 It was first proposed by Sivers \cite{Sivers1990} that it may be possible to 
explain this asymmetry by allowing  a correlation between 
the transverse momentum of the quark and the polarization of parent hadron.
This approach requires a pQCD factorization scheme that includes the  spin and intrinsic transverse momentum effects. 
With  the inclusion of ${\bkt}$ dependence in parton distribution functions (pdf's) and fragmentation functions (ff's)
\cite{tmd-fact1}, one is led  to a generalized factorization formula called transverse momentum dependent TMD factorization
 \cite{Sivers1990, tmd-fact2}.
TMD factorization in some processes has been 
proved at leading twist and leading order \cite{fact} and has been argued to hold 
at all orders.  

The inclusion of the effect of transverse momentum of partons in pdf's and ff's
leads to a new class of parton distributions 
which are obtained as extensions of usual collinear pdf's and include the effects of spin and transverse momentum 
of the partons. One of these functions is the  Sivers function, which describes the probability of finding an 
unpolarized parton inside a transversely polarized hadron. The coupling of the transverse 
momentum of the unpolarized quarks and gluons to the nucleon spin is in fact related to 
their orbital angular momentum. Thus the Sivers asymmetry gives access to the 
orbital angular momentum of the partons. The number density of  partons  
inside a proton with transverse polarization ${\bf S}$ and momentum $\bfP$ is parameterized as \cite{Anselmino2003-PRD67}
\be
 f_{a/\pup}(x_a,\bkt_a,\bfS) \equiv
f_{a/p}(x_a,{\kt}_a) + \frac{1}{2}\Delta^Nf_{a/\pup}(x_a,{\kt}_a)
\> \hat{\bfS}\cdot (\hat{\bfP}\times \hat{\bfk}_{\perp a})
\label{parm-siv}
\ee
where ${\bkt}_a$ is the transverse momentum of the parton, $x_a$ is the longitudinal momentum fraction of parton,
$f_{a/p}(x, {\bkt}_a)$ is the unpolarized parton distribution and $\Delta^Nf_{a/\pup}(x,k_{\perp a})$ is the Sivers function.
In this work, we propose charmonium production as a probe to investigate the 
Sivers function and as a first step, estimate SSA in photoproduction of charmonium in scattering of 
electrons off transversely polarized protons. 

One of the difficulties in getting information about the spin and transverse momentum dependent pdf's and 
fragmentation functions is that very often two or more of these functions contribute to the same physical observable 
making it difficult to estimate each single one separately. 
It has been shown  how properly defined SSA's in the Drell-Yan process depend only on the quark Sivers function and the 
unpolarized quark distributions \cite{Anselmino2003-PRD67}. The studies of Anselmino $\it{etal}$  show that the  magnitude 
of the Sivers asymmetry in the Drell-Yan (DY) process for forthcoming experiments at RHIC, COMPASS, J-PARC, PAX, PANDA, NICA and SPASCHARM  is 
large \cite{Anselmino2009}. A  study of the Sivers effect for pion and kaon production in 
semi-inclusive deep inelastic scattering (SIDIS) processes has been  performed and estimates have been given for experiments at COMPASS  
and JLab \cite{Anselmino-PRD72,kp09, {Anselmino:2005nn}}. 
It has been proposed to probe the gluon Sivers function by looking at back-to-back correlations in $p^\uparrow p $ interactions at 
RHIC \cite{Boer-PRD69(2004)094025}. Another process that has been suggested as
a probe to access gluon Sivers function is $p^\uparrow p \rightarrow D X$ as the SSA in this case arises due to the 
gluon Sivers function alone mainly in the intermediate rapidity region \cite{Anselmino2004}.

In this paper, we investigate feasibility of using charmonium production to obtain information about the Sivers function.
Charmonium production has been known to be a sensitive tool to  study QCD for bound states of heavy
quark-antiquark systems. Recently, there has been some discussion about the possibility to use fixed target experiments at LHC
for charmonium production with the aim of investigating the quarkonium production mechanism~\cite{kurepin-qm2011}. 
Photoproduction and electroproduction of charmonium near threshold are 
expected to throw some light  on this mechanism as well as on hadron 
structure \cite{hoyer1997}. In fact the connection between charm, charmonium
production  and the gluon densities has been explored since long  for
protons-polarized and unpolarized-, nuclei and  photons~\cite{jpsi-gluon}. 
Here, we study asymmetry in photoproduction (i.e. 
low virtuality electroproduction) of charmonium in scattering off polarized 
protons. At LO, this  receives contribution only from a single partonic 
subprocess $\gamma g \rightarrow c {\bar c}$ . Hence,  SSA in 
$e + p^\uparrow \rightarrow e +J/\psi +X$, if observed, can be used as a clean 
probe of gluon Sivers function. In addition, charmonium production mechanism 
can also have implications for this SSA and therefore, its study can help 
probe the production mechanism for charmonium.  

There are three models for charmonium production. In the color singlet model \cite{sing} 
the cross section for charmonium production is factorized into a short distance part for 
$c {\bar c}$ pair production 
calculable in perturbation theory and a nonperturbative matrix element
for the formation of a bound state, which is produced in a color singlet state.
In color evaporation model, first proposed by Halzen and Matsuda 
\cite{hal} and Fritsch \cite{fri} a statistical treatment of color is made and 
the probability of finding a specific quarkonium state is assumed to be independent 
of the color of heavy quark pair. In later versions of this model it has been found
that the data are better fitted if a phenomenological factor is included in the differential 
cross section formula, which depends on a Gaussian distribution of the transverse 
momentum of the charmonium \cite{ce2}.
A more recent model of charmonium production is the color octet model \cite{octet}.
This is based on a factorization approach in nonrelativistic QCD, and it allows $c {\bar c}$ pairs   
to be produced in color octet states. Here again, one 
requires knowledge of the nonperturbative color octet matrix elements,
which are determined  through fits to the data on charmonium production. 
Inclusion of nonzero intrinsic transverse momentum for the colliding partons
can help us to understand the discrepancy between these  matrix elements 
determined from the hadroproduction (Tevatron) data and the leptoproduction 
(HERA) data~\cite{sridhar-kt, hagler-kt}. 
Since the nonrelativistic QCD calculations are done in  collinear approximation, it is not surprising that the effects 
of nonzero transverse momentum of the colliding partons can be large.
In this work, we have chosen to work with the color evaporation model (CEM) as its simplicity makes it 
suitable for an initial study of SSA in the charmonium production.      

It has been proposed that the SSA's at leading twist in pQCD, arise from the final state interactions 
between the outgoing quark and the target spectator system \cite{ssa-twist}. In case of the Sivers asymmetry,
the initial and/or final state interactions can generate a nonzero phase in the amplitude, 
which through the naive time reversal odd (T-odd) Sivers function then gives rise to the SSA. This interaction is part of the 
gauge link  present in the TMD functions and depends on the specific process
under consideration. Thus it introduces a process dependence, in particular, in the T-odd distribution 
functions \cite{pro}. In fact the direction of the link is opposite in the
Drell-Yan process as compared to SIDIS. For a generic hadronic process, it
can be completely different from the above two processes. The generation of SSA in $J/\psi$ production
in $ep^\uparrow$ and $p p^\uparrow$ was studied in \cite{feng}. It was found therein that existence of
nonzero SSA depends on the production mechanism of $J/\psi$, namely in  $ep^\uparrow$ processes nonzero 
SSA is expected only if the charmonium is produced in the color octet state and for $p p^\uparrow$  it should be produced in the 
color singlet state. However no numerical estimate of the asymmetry was given.

We provide a first estimate of the SSA in the low $Q^2$ electroproduction (photoproduction) 
of $J/\psi$ at leading order, using a factorized formula along with models of Sivers function used in the literature. 
As a first step in our investigation of SSA's  in charmonium production, 
we have used the color evaporation model for charmonium production.  
According to CEM, the cross section for charmonium production is proportional to the
rate of production of $c\bar{c}$ pair integrated over the mass range $2m_c$ to $2m_D$ \cite{cem0}
\be
\sigma=\frac{1}{9}\int_{2m_c}^{2m_D} dM_{c\bar{c}} \frac{d\sigma_{c\bar{c}}}{dM_{c\bar{c}}}
\label{sigmacem}
\ee  
where $m_c$ is the charm quark mass and $2m_D$ is the $D\bar{D}$ threshold.

At leading order (LO) there are no ff's involved  and the only contribution to asymmetry comes from the Sivers function.
Therefore, we can use this observable to extract information about it.

Here, we have used Weizsacker-Williams equivalent photon approximation to 
calculate cross section for the process 
$e+\pup\rightarrow e+ J/\psi+X$. The underlying  partonic process at 
LO is $\gamma g\rightarrow c\bar{c}$ and therefore,
the only $\kt$ dependent pdf appearing is the gluon Sivers function.
For a complete calculation of photoproduction of $J/\psi$ one has to consider 
higher order contributions and also the resolved photon  contributions \cite{ce2}.

To assess  numerical estimates, we  have used and compared  two models  of the quark Sivers function 
obtained from SIDIS data and have estimated the magnitude of asymmetry at JLab, HERMES, COMPASS and eRHIC center of mass energies. 
We predict nonzero asymmetry in both rapidity and ${q}_T$ distribution for the parameter set fitted from 
experimental data \cite{2011-parmeterization}.

\section{FORMALISM FOR ASYMMETRY IN $J/\psi$ PRODUCTION}

\subsection{Color Evaporation Model}

We consider the LO parton model cross section for low virtuality electroproduction (photoproduction)
of $J/\psi$ within color evaporation model. According to CEM, the cross section for charmonium production 
is given by Eq. (\ref{sigmacem}), where  $\displaystyle\frac{d\hat{\sigma}_{c\bar{c}}}{dM_{c\bar{c}}}$ is calculable perturbatively, 
$M_{c\bar{c}}$ being the invariant mass of the $c\bar{c}$ pair.
The differential cross section for the $\gamma+p\rightarrow J/\psi+X$ is given by
\be
\frac{d\sigma^{\gamma p\rightarrow c\bar{c}}}{dM^2_{c\bar{c}}}=
\int dx f_{g/p}(x) \frac{d\hat{\sigma}^{\gamma g\rightarrow c\bar{c}}}{dM^2_{c\bar{c}}}
\label{xsec-ccbar}
\ee
where $f_{g/p}(x)$ is the gluon distribution in the proton. 

Using the Weizsacker-Williams approximation \cite{wwf1,wwf2}, one can convolute the 
cross section given by Eq. (\ref{xsec-ccbar}) with a photon flux factor to obtain the electroproduction cross section 
for $e+p\rightarrow e+ J/\psi+X$ 
\be
\frac{d\sigma^{ep\rightarrow e+ J/\psi+X}}{d M_{c\bar{c}}^2}=\int dy\> f_{\gamma/e}(y)\> 
\frac{d\hat{\sigma}^{\gamma p\rightarrow c\bar{c}}}{d M_{c\bar{c}}^2}
\label{xsec-ep}
\ee
where $y$ is the energy fraction of electron carried by the photon and $f_{\gamma/e}(y)$
is the distribution function of the photon in the electron given by \cite{ww},
\bea
f_{\gamma/e}(y,E)=\frac{\alpha}{\pi} \{\frac{1+(1-y)^2}{y}\left(ln\frac{E}{m}-\frac{1}{2}\right)
+\frac{y}{2}\left[ln\left(\frac{2}{y}-2\right)+1\right] \nonumber \\
+\frac{(2-y)^2}{2y}ln\left(\frac{2-2y}{2-y}\right) \}.
\label{ww-function}
\eea

Thus, the cross section for electroproduction of $J/\psi$ using  WW approximation is given by,
\be
\sigma^{ep\rightarrow e+J/\psi+X}=
\int_{4m_c^2}^{4m_D^2} dM^2_{c\bar{c}} \int dy\> dx\> f_{\gamma/e} (y)\> f_{g/p}(x) 
\>\frac{d\hat{\sigma}^{\gamma g\rightarrow c\bar{c}}}{dM_{c\bar{c}}^2}.
\label{xsec-gammap}
\ee

\subsection{Single spin asymmetry in $J/\psi$ production}

To calculate SSA in scattering of electrons off a polarized proton target, we assume
a generalization of CEM expression by taking into account the transverse momentum dependence
of the Weizsacker-Williams function and gluon distribution function 
\be
\frac{d\sigma^{e+p^\uparrow\rightarrow e+J/\psi + X}}{d M^2}=
\int  dx_\gamma\> dx_g\> [d^2\bfk_{\perp\gamma}d^2\bfk_{\perp g}]\>
f_{g/p^{\uparrow}}(x_{g},\bfk_{\perp g})
f_{\gamma/e}(x_{\gamma},\bfk_{\perp\gamma})
\frac{d\hat{\sigma}^{\gamma g\rightarrow c\bar{c}}}{dM^2}
\label{dxec-ep}
\ee
where $M^2\equiv M_{c\bar{c}}^2$. We have not written the scale dependence
of the quantities on the {\it right hand side} explicitly. 
As mentioned earlier, a generalization of factorization formula
involving TMD pdf's and ff's leads to nonzero SSA in DY, SIDIS and other processes. Therefore, we expect 
that inclusion of transverse momentum dependence in WW function and generalization of the CEM 
expression might be a valid approach to estimate SSA in charmonium production. 

SSA is defined as in Eq. (1), 
where in our case $d\sigma ^{\uparrow(\downarrow)}$ are single transverse spin-dependent cross sections for 
$e+p^\uparrow + \rightarrow e+ J/\psi+X$ and $e+p^\downarrow \rightarrow e + J/\psi+X$, respectively.

The difference in $d\sigma^\uparrow$ and $d\sigma^\downarrow$ is parameterized in terms of the gluon
Sivers function
\be
d\sigma^\uparrow-d\sigma^\downarrow=
\int dx_\gamma\> dx_g\> d^2\bfk_{\perp\gamma}\>d^2\bfk_{\perp g}\>
\Delta^{N}f_{g/p^{\uparrow}}(x_{g},\bfk_{\perp g})\>
f_{\gamma/e}(x_{\gamma},\bfk_{\perp\gamma})\>
{d\hat{\sigma}}^{\gamma g\rightarrow c\bar{c}}
\label{dxsec-nssa}
\ee
where $d\hat{\sigma}$ is the elementary cross section for the process $\gamma g\rightarrow c\bar{c}$
given by

\be
d\hat\sigma = \frac{1}{2\hat s} \> \frac{d^3p_c}{2E_c} \frac{d^3p_{\bar{c}}}{2E_{\bar{c}}}\>  
\frac{1}{(2\pi)^2} \> \delta^4(p_\gamma + p_g - p_c - p_{\bar{c}}) \>
\overline{\left\vert \, M_{\gamma g \to c\bar{c}} \, \right\vert^2} \>. 
\label{ecs}
\ee

We rewrite $\displaystyle\frac{d^3p_{\bar{c}}}{2E_{\bar{c}}}=d^4p_{\bar{c}}\>
\delta({p_{\bar{c}}}^2-m^2_c)$
and change the variable to $q=p_c+p_{\bar{c}}$
\cite{Anselmino2003-PRD67}  so that 
\be
\frac{d^3p_{\bar{c}}}{2E_{\bar{c}}}=d^4q \> \delta((q-p_{c})^2-m^2_c).
\label{phase-space}
\ee

Now using the expression for total partonic cross section 
\be
\hat{\sigma_0}^{\gamma g\rightarrow c\bar{c}}(M^2)= \frac{1}{2\hat s} \int  \frac{d^3p_c}{2E_c} \>  
\frac{1}{(2\pi)^2} \> \delta((q-p_{c})^2-m^2_c) \>
\overline{\left\vert \, M_{\gamma g \to c\bar{c}} \, \right\vert^2} \>
\label{tot-xsec}
\ee
and changing the variables from $q_0$ and $q_L$ to $M^2$ and rapidity $y$ so that 
\be
dM^2 dy=2 dq_0 dq_L
\ee
We finally obtain
\bea
\frac{d^{4}\sigma^\uparrow}{dydM^{2}d^2\bfq_T}-\frac{d^4\sigma^\downarrow}{dydM^{2}d^2\bfq_T}=
\frac{1}{2}\int [dx_{\gamma}d^2\bfk_{\perp\gamma} dx_{g} d^2\bfk_{\perp g}]
\Delta^{N}f_{g/p^{\uparrow}}(x_{g},\bfk_{\perp g}) \nonumber \\ 
\times f_{\gamma/e}(x_{\gamma},\bfk_{\perp\gamma}) 
\delta^{4}(p_{g}+p_{\gamma}-q)\>
\hat\sigma_{0}^{\gamma g\rightarrow c\bar{c}}(M^2). 
\label{nssa}
\eea
where the partonic cross section is given by \cite{gr78}
\be
\hat{\sigma_0}^{\gamma g\rightarrow c\bar{c}}(M^2)=
\frac{1}{2}e_{c}^2\frac{4\pi\alpha\alpha_s}{M^2}
[(1+\gamma-\frac{1}{2}\gamma^2)\ln{\frac{1+\sqrt{1-\gamma}}{1-\sqrt{1-\gamma}}}
-(1+\gamma)\sqrt{1-\gamma}].
\label{jpsics}
\ee
Here, $\displaystyle\gamma=\frac{4 m_c^2}{M^2}$ and $M^2\equiv\hat{s}$.

$\Delta^{N}f_{g/p^{\uparrow}}(x_{g},\bfk_{\perp g})$  is related to the gluon Sivers function
$\Delta^N f_{g/p^{\uparrow}}(x, k_{\perp g})$ by 
\be
 \Delta^{N}f_{g/p^{\uparrow}}(x_{g},\bfk_{\perp g})= \Delta^N f_{g/p^{\uparrow}}(x_g, k_{\perp}) \> \hat{\bfS} \cdot
(\hat{\bfP} \times \hat{\bfk}_{\perp g}).\>
\label{delf2} 
\ee
Similarly the total cross section is given by
\bea
\frac{d^{4}\sigma^\uparrow}{dydM^{2}d^2\bfq_T}+\frac{d^{4}\sigma^\downarrow}{dydM^{2}d^2\bfq_T}=
\int [dx_{\gamma}d^2\bfk_{\perp\gamma} dx_{g} d^2\bfk_{\perp g}]
f_{g/p}(x_{g},\bfk_{\perp g}) \nonumber \\
\times f_{\gamma/e}(x_{\gamma},\bfk_{\perp\gamma}) 
\delta^{4}(p_{g}+p_{\gamma}-q)
\hat\sigma_{0}^{\gamma g\rightarrow c\bar{c}}(M^2). 
\label{dssa}
\eea
It is worth mentioning at this point that following the procedure used by Anselmino etal 
in case of Drell-Yan process, we have been able to write the distribution in $M^2$, $q_T$ and y of the 
produced $J/\psi$ in terms of unpolarized total partonic cross section. 

Rewriting the four momentum conservation $\delta$ function as
\bea
\delta^{4}(p_{g}+p_{\gamma}-q)&=&
\delta(E_g + E_\gamma - q_0) \, \delta(p_{z_g} + p_{z_\gamma} - q_L)
\delta^2(\bfk_{\perp\gamma}+\bfk_{\perp g}-\bfq_T) \nonumber\\
&=& \frac{2}{s} \, \delta \! \left( x_g - \frac{M}{\sqrt s} \, e^y \right) \, 
\delta \! \left( x_\gamma - \frac{M}{\sqrt s} \, e^{-y} \right)
\delta^2(\bfk_{\perp\gamma}+\bfk_{\perp g}-\bfq_T) \label{delta-4}
\eea
one can perform the $x_\gamma$ and $x_g$ integrations
to obtain
\bea
\frac{d^{4}\sigma^\uparrow}{dydM^2d^2\bfq_T}-\frac{d^4\sigma^\downarrow}{dydM^2d^2\bfq_T}=
\frac{1}{s}\int [d^2\bfk_{\perp\gamma}d^2\bfk_{\perp g}]
\Delta^{N}f_{g/p^{\uparrow}}(x_{g},\bfk_{\perp g})
f_{\gamma/e}(x_{\gamma},\bfk_{\perp\gamma}) \nonumber\\
\times\>\delta^2(\bfk_{\perp\gamma}+\bfk_{\perp g}-\bfq_T)
\hat\sigma_{0}^{\gamma g\rightarrow c\bar{c}}(M^2)
\label{num-ssa}
\eea

and
\bea
\frac{d^{4}\sigma^\uparrow}{dydM^2d^2\bfq_T}+\frac{d^4\sigma^\downarrow}{dydM^2d^2\bfq_T}=
\frac{2}{s}\int [d^2\bfk_{\perp\gamma}d^{2}\bfk_{\perp g}]
f_{g/p}(x_g,\bfk_{\perp g})
f_{\gamma/e}(x_{\gamma},\bfk_{\perp\gamma}) \nonumber\\
\times\>\delta^2(\bfk_{\perp\gamma}+\bfk_{\perp g}-\bfq_T)
\hat\sigma_{0}^{\gamma g\rightarrow c\bar{c}}(M^2)
\label{den-ssa}
\eea

with 
\be
x_{g,\gamma} = \frac{M}{\sqrt s} \, e^{\pm y}. 
\label{x-gammag}
\ee
Integrating Eqs. (\ref{num-ssa})  and (\ref{den-ssa}) over $M^2$ as prescribed by CEM, 
we obtain  the difference and sum of $\displaystyle\frac{d^{3}\sigma^\uparrow}{dyd^2\bfq_T}$ 
and $\displaystyle\frac{d^3\sigma^\downarrow}{dyd^2\bfq_T} $for $J/\psi$ production. 

We follow the convention  in Ref.~\cite{vogelsang-weight} and define the Sivers asymmetry integrated 
over the azimuthal angle of $J/\psi$ with a weight factor  $\sin({\phi}_{q_T}-\phi_S)$ :
\be
A_N^{\sin({\phi}_{q_T}-\phi_S)} =\frac{\int d\phi_{q_T}
[d\sigma ^\uparrow \, - \, d\sigma ^\downarrow]\sin({\phi}_{q_T}-\phi_S)}
{\int d{\phi}_{q_T}[d{\sigma}^{\uparrow} \, + \, d{\sigma}^{\downarrow}]}
\label{weight-ssa}
 \ee
where $d\sigma^\uparrow$ is differential cross section in $q_T$ or y variable and 
${\phi}_{q_T} $ and $\phi_S$ are the azimuthal angles of the $J/\psi$ and proton spin respectively. 
To evaluate asymmetry in $y$ distribution, we will substitute 
\bea
d\sigma ^\uparrow \, - \, d\sigma ^\downarrow=\int d\phi_{q_T}\int q_T\>dq_T
\int_{4m^2_c}^{4m^2_D}[dM^{2}]\int[d^2\bfk_{\perp g}]
\Delta^{N}f_{g/p^{\uparrow}}(x_{g},\bfk_{\perp g}) \nonumber \\
\times\>f_{\gamma/e}(x_{\gamma},\bfq_T-\bfk_{\perp g}) \>
\hat\sigma_{0}(M^2)\>\sin({\phi}_{q_T}-\phi_S) \label{dxsec-y}
\eea
and
\bea
d\sigma ^\uparrow \, + \, d\sigma^\downarrow= 
2\int d\phi_{q_T}\int q_T\>dq_T\int_{4m^2_c}^{4m^2_D}[dM^{2}]\int[d^{2}\bfk_{\perp g}]
f_{g/p}(x_g,\bfk_{\perp g}) \nonumber \\
\times\>f_{\gamma/e}(x_{\gamma},\bfq_T-\bfk_{\perp g})\,
\hat{\sigma}_0(M^2). 
\label{txsec-y}
\eea

Thus at LO, the SSA depends on Weizsacker-Williams function, gluon distribution function 
and gluon Sivers function. We discuss our choice of WW function and Sivers function in the 
following subsection.

\subsection{Sivers function and Weizsacker-Williams function}

In our analysis, we have chosen a  kinematical configuration in which proton with momentum ${\bfP}$
is moving along z axis and is transversely polarized 
in y direction so that 
\be
\hat{\bfS}\cdot (\hat{\bfP}\times \hat{\bfk}_{\perp g})={\hat k}_{\perp g z}=\cos{{\phi}_{\kt}}
\label{phikt}
\ee
where, $\bfk_{\perp g} = k_{\perp}(\cos\phi_{k_\perp}, \, \sin\phi_{k_\perp}, \, 0).$

For $k_{\perp g}$ dependence of the unpolarized pdf's, we use a 
simple factorized and Gaussian form \cite{Anselmino-PRD72}
\be
f_{g/p}(x_g,k_{\bot g})=f_{g/p}(x_g)\frac{1}{\pi\langle k^{2}_{\bot g}\rangle} 
e^{-k^{2}_{\bot g}/\langle{k^{2}_{\bot g}\rangle}}.
\label{gauss}
\ee

We also need a transverse momentum dependent WW function. 
In our analysis, we have used two choices for it:
\begin{itemize}
\item[1)] A simple Gaussian form as above : 
\be
f_{\gamma/e}(x_\gamma,k_{\bot \gamma})=f_{\gamma/e}(x_\gamma)\frac{1}{\pi\langle k^{2}_{\bot \gamma}\rangle} 
e^{-k^{2}_{\bot \gamma}/\langle{k^{2}_{\bot \gamma}\rangle}}.
\label{gauss-g}
\ee 

\item[2)] A dipole form :
\be
f_{\gamma/e}(x_\gamma,k_{\bot \gamma})=
f_{\gamma/e}(x_\gamma)\>\frac{1}{2\pi}\>\frac{N}{k_{\perp \gamma}^2+{k_0}^2}
\label{dipole}
\ee
where $N$ is a normalization constant, which gets cancelled in the asymmetry.
\end{itemize}

For Sivers function we have used two models in our analysis
\begin{itemize}

\item In most part of the analysis we use model I 
\cite{Anselmino2009} which is given by,  
\be
\Delta^{N}f_{g/p^{\uparrow}}(x_g,{\bfk_{\perp}}_g)=
\Delta^{N}f_{g/p^{\uparrow}}(x_g)\frac{1}{\pi\langle {k}_{\perp g}^2\rangle}\> 
h({\kt}_g)\>e^{-{k}_{\perp g}^2/\langle {k}_{\perp g}^2\rangle}\cos({\phi}_{k_{\perp}})
\label{sivers2}
\ee
where the gluon Sivers function, $\Delta^Nf_{g/p^{\uparrow}}(x_g)$ is defined as 
\be
\Delta^Nf_{g/p^{\uparrow}}(x_g) = 2\,{\mathcal N}_g(x_g)\,f_{g/p}(x_g)\,
\label{dnf}
\ee
where ${\mathcal N}_g(x_g)$ is an x-dependent normalization for gluon to be chosen so that the 
gluon Sivers function obeys the  positivity bound
\be
\frac{|\Delta^N f_{g/p^{\uparrow}}(x_g, {\bkt}_g)|}
{2 \,\hat f_{g/p}(x_g, {\kt}_g)}  \leq 1,
\quad\quad \forall\, x_g,\,{\kt}_g \,
\label{posb}
\ee 
and \be
h({\kt}_g) = \sqrt{2e}\,\frac{{\kt}_g}{M_{1}}\,e^{-{k}_{\perp g}^2/{M_{1}^2}}\> 
\label{siverskt}
\ee
where the gluon Sivers function, $\Delta^Nf_{g/p^{\uparrow}}(x_g)$ is given as in model I 
and $M_1$ is parameter obtained by fitting the recent experimental data corresponding to pion and kaon 
production at HERMES and COMPASS. 

The corresponding parameterizations for quark Sivers function ${\mathcal N}_u(x)$ 
and ${\mathcal N}_d(x)$ have been 
fitted from SIDIS data and are given by \cite{Boer-PRD69(2004)094025},
\be 
{\mathcal N}_f(x) = N_f x^{a_f} (1-x)^{b_f} \frac{(a_f + b_f)^{(a_f +
b_f)}}{{a_f}^{a_f} {b_f}^{b_f}} \; .
\label{admfct}
\ee
where $a_f, b_f, N_f$ for u and d quarks are free parameters 
obtained by fitting the data. However, there is no information available on ${\mathcal N}_g(x)$.
In our analysis we have used two choices \cite{Boer-PRD69(2004)094025}
\begin{itemize}
\item[(a)] ${\mathcal N}_g(x)=\left( {\mathcal N}_u(x)+
{\mathcal N}_d(x) \right)/2 \;$,
\item[(b)] ${\mathcal N}_g(x)={\mathcal N}_d(x)$.
\end{itemize}

\item We also compared the results with  model II~ \cite{Anselmino:2005nn, Anselmino2004}  given by  
\be
\Delta^{N}f_{g/p^{\uparrow}}(x_g,{\bkt}_g)=
\Delta^{N}f_{g/p^{\uparrow}}(x_g)\frac{1}{\pi\langle{k}_{\perp g}^2\rangle} 
e^{-{\kt^2}_g/\langle{\kt^2}_g\rangle}
\frac{2{\kt}_g M_0}{{\kt^2}_g + M_0^{2}}\cos(\phi_{\kt}), 
\label{siv}
\ee\\
and $M_0=\sqrt{\langle{{k}_{\perp g}^2}\rangle}$
where the gluon Sivers function is given as in model I. 
 
\end{itemize}
 
Model I has been used in analysis of SSA in the SIDIS \cite{Anselmino-PRD72}
 and DY processes \cite{Anselmino2009} and model II has been used for the quark
 Sivers function to estimate SSA in D meson production at RHIC \cite{Anselmino2004}. 
 A comparison of $k_\perp$ dependence of Sivers function in the two models is given by Fig.~7 of 
 Ref.~\cite{Anselmino:2005nn}. 
We will give numerical estimates of   asymmetry in photoproduction of charmonium using both of 
these  models and will also 
compare both the parameterizations for ${\mathcal N}_g(x_g)$ in our analysis in 
the next section.

\section{NUMERICAL ESTIMATES  FOR THE ASYMMETRY IN $J/\psi$ PRODUCTION}

We will now estimate the magnitude of asymmetry using models I and II for both the 
parameterizations (a) and (b) and Gaussian form for WW function $\kt$ -dependence. 

The values of best fit parameters of Sivers functions we have used are
\cite{2011-parmeterization}
\bea
N_u = 0.40, \ a_u=0.35, \ b_u =2.6 \; , \nonumber \\
N_d = -0.97, \ a_d = 0.44, \ b_d=0.90 \;, \nonumber \\
M_1^2=0.19~GeV^2.
\label{2011-parm}
\eea
These parameters are from new HERMES and COMPASS data \cite{hermes09,compass09} 
fitted at $<Q^2> = 2.4 GeV^2$. 

The value of $\langle{k_{\perp g}^2}\rangle$ is chosen to be the  same 
as $\langle{k_{\perp}^2}\rangle $ for 
quarks obtained in Ref.~\cite{Anselmino:2005nn} by analysis 
of Cahn effect in unpolarized SIDIS 
from data collected in different energy and $Q^2$ ranges assuming a 
constant Gaussian width. The value of 
$\langle{k_{\perp \gamma}^2}\rangle$ has been chosen to be comparable to  $\langle{k_{\perp g}^2}\rangle$. 
As pointed out in \cite{apana}, $\langle{k_{\perp}^2}\rangle$ in fact depends on energy. 
However, in this paper, we keep this value fixed as above as we do not expect a large variation 
in this parameter related to charmonium production in the model concerned (see below).   

\subsection{Asymmetry using model I}

In model I, the exponential nature of the function $h(\kt)$ enables one to perform the $k_\perp$ 
integration analytically and we obtain the expressions for numerator and denominator in asymmetry as 
\bea
\frac{d^{3}\sigma^\uparrow}{dyd^2\bfq_T}-\frac{d^{3}\sigma^\downarrow}{dyd^2\bfq_T}
=\frac{1}{s}\int_{4m^2_c}^{4m^2_D}dM^2
\Delta^{N} f_{g/p^\uparrow}(x_{g})f_{\gamma/e}(x_{\gamma})
\sqrt{2e}\frac{q_T}{M_1}\nonumber\\
\times\>\frac{\langle{k_{S}^2}\rangle^2 \exp[-q_T^2/(\langle{k_{S}^2}\rangle+
\langle{{k_{\perp\gamma}^2}}\rangle)]
}{\pi[\langle{k_{S}^2}\rangle+\langle{{k}_{\perp\gamma}^2}\rangle]^2
 \langle{{k}_{\perp g}^2}\rangle}
\cos({\phi}_{q_T})\>\hat\sigma_{0}^{\gamma g\rightarrow c\bar{c}}(M^2) 
\label{num2}
\eea

and,
\bea
\frac{d^{3}\sigma^\uparrow}{dyd^2\bfq_T}+\frac{d^{3}\sigma^\downarrow}{dyd^2\bfq_T}
=\frac{2}{s}\int_{4m^2_c}^{4m^2_D}dM^2
f_{g/p}(x_{g})f_{\gamma/e}(x_{\gamma}) \nonumber\\
\times\>\frac{\exp[-q_T^2/(\langle{{k}_{\perp g}^2}\rangle+\langle{{k}_{\perp\gamma}^2}\rangle)]
}{\pi[\langle{{k}_{\perp g}^2}\rangle+\langle{{k}_{\perp\gamma}^2}\rangle]}
\>\hat\sigma_{0}^{\gamma g\rightarrow c\bar{c}}(M^2) \label{den-m1}
\eea
where 
\be
\frac{1}{\langle k_{S}^2\rangle} =
\frac{1}{M_1^2}+\frac{1}{\langle k_{\perp g}^2\rangle}\> .
\label{ksq}
\ee


\subsection{Asymmetry using Model II}

With the $k_\perp$ dependent unpolarized pdf's and gluon Sivers function 
in Eq.~(\ref{gauss}) and (\ref{siv}) numerator becomes
\bea
\frac{d^{3}\sigma^\uparrow}{dy\>d^2\bfq_T}-\frac{d^{3}\sigma^\downarrow}{dy\>d^2\bfq_T}
=\frac{1}{s}\int_{4m^2_c}^{4m^2_D}dM^2\int_{0}^{{{\kt}_g}_{max}}d{\kt}_g
\int_{0}^{2\pi}d\phi_{k_\perp}{\kt}_g\>
\Delta^{N} f_{g/p^\uparrow}(x_{g})\>f_{\gamma/e}(x_{\gamma})\nonumber \\
\times\>\exp[{-{k}_{\perp g}^2/\alpha}]\>\exp[{-{k}_{\perp g}^2/\beta}]\>\exp[{-q^2_T/\alpha}]\>
\exp[{(2q_T\>{\kt}_g \cos(\phi_{k_\perp}-\phi_{q_T}))/\alpha}] \nonumber \\
\times\>\frac{1}{\pi^2\>\alpha\>\beta}\>\frac{2 {\kt}_g M_0}{{k}_{\perp g}^2+M_0^{2}}\>
\cos(\phi_{k_\perp})\>\hat\sigma_{0}^{\gamma g\rightarrow c\bar{c}}(M^2) 
\label{num1}
\eea
and the denominator becomes,
\bea
\frac{d^{3}\sigma^\uparrow}{dy\>d^2\bfq_T}+\frac{d^{3}\sigma^\downarrow}{dy\>d^2\bfq_T}
=\frac{2}{s}\int_{4m^2_c}^{4m^2_D}dM^2\int_{0}^{{{\kt}_g}_{max}}d{\kt}_g
\int_{0}^{2\pi}d\phi_{k_\perp}{\kt}_g\>
f_{g/p}(x_{g})\>f_{\gamma/e}(x_{\gamma}) \nonumber \\
\times\>\exp[{-{k}_{\perp g}^2/\alpha}]\>\exp[{-{k}_{\perp g}^2/\beta}]\>\exp[{-q^2_T/\alpha}]\>
\exp[{(2\>q_T\>{\kt}_g\>\cos(\phi_{k_\perp}-\phi_{q_T}))/\alpha}]\nonumber \\
\times\>\frac{1}{\pi^2\>\alpha \> \beta}\>
\hat\sigma_{0}^{\gamma g\rightarrow c\bar{c}}(M^2)
\label{den1}
\eea
where $\alpha=\langle(\bfq_T-{\bkt}_g)^2\rangle$
, $\beta=\langle {{k}_{\perp g}^2}\rangle$ and $M_0=\sqrt{\langle {k}_{\perp g}^2\rangle}=\sqrt{\beta}$.

We have used ${{\kt}_g}_{max} = 0.5$ GeV, observing that the effective intrinsic motion is limited to 
$\kt\leq\sqrt{\langle{\kt^2}\rangle}$ for Gaussian distribution \cite{kp09}.

We have estimated the asymmetry with both kinds of parameterizations 
[labeled (a) and (b)].
The estimates are obtained using GRV98LO for gluon distribution functions and 
Weizsaker-Williams function for photon distribution \cite{ww}. 
The scale of the gluon distribution 
 has been taken to be $ \hat s (= M^2)$~ \cite{gr78}. It is worth
pointing out that the scale evolution of the TMD's including the Sivers
function has been worked out in \cite{evo1,evo2} and recently it 
has been noted in \cite{evo3,evo4} that in SIDIS the evolution indeed 
affects the Sivers asymmetry. In fact the evolution effectively produces 
a change in the Gaussian width of the TMDs depending on the scale. The possible effect of such scale dependence 
is not included in our this, exploratory, study of  the asymmetry in $J/\psi$
production and is deferred to a later publication. However, since we are 
using CEM at LO, the hard scale is between $4m_c^2$ and $4m_D^2$ independent 
of the beam energies and the TMD evolution effects within this rather narrow 
range will be similar for all experiments considered. 
Our results are presented in 
Figs. (1)-(9). The main features of these plots are summarized below. 

In Figs. (1)-(5) we have shown the asymmetry ($A_N^{\sin({\phi}_{q_T}-\phi_S)}$)
as a function of rapidity $y$ and $q_T$ 
respectively for JLab ($\sqrt s = 4.7$ GeV), HERMES ($\sqrt s = 7.2 $ GeV), 
COMPASS ($\sqrt s = 17.33$ GeV) and eRHIC ($\sqrt s = 31.6 $ GeV and
$\sqrt s = 158.1 $ GeV) energies.  
The plots are for models I(a) and I(b) for two choices of the gluon Sivers function. 
We obtain sizable asymmetry in the kinematical regions of all the experiments 
for model I (b). The asymmetry is smaller in model I(a). The asymmetry decreases 
with increase of $y$ for JLab. At HERMES it remains almost independent of $y$
for negative rapidity and decreases with increasing $y$ for positive rapidity.
For COMPASS as well as for eRHIC lower energy the asymmetry increases with $y$,
reaches a maximum and then decreases. This maximum is reached at $y \approx 0.6$ for 
COMPASS and at $y \approx 1.2$ for eRHIC for model I(b). For the higher energy at eRHIC 
no maximum is seen in the rapidity region plotted. The asymmetry increases with the 
increase of $y$. Both model I(a) and model  I(b) show similar qualitative behavior.
The asymmetry increases with $q_T$ for both models, and for higher values of 
$q_T \approx 0.6-0.7$ GeV, it becomes relatively steady.  

In Figs. 6 and 7 we have compared the results for the asymmetry for models I and II at COMPASS energy 
($\sqrt s = 17.33$ GeV). Figure 6 shows the asymmetry for models I(a) and II(a) and 
Fig. 7 shows the same for models I(b) and II(b). In model II the asymmetry is larger 
in magnitude. It increases steadily with $q_T$ in model II unlike in model I
where it first increases rapidly then becomes relatively stable with increase of $q_T$.
The difference in the $q_T$ dependence in model I and model II arises due to the
different $k_\perp$ dependence of the Sivers function in the two models.
Figure 8 shows a comparison of the $y$ and $q_T$ dependence of the asymmetry 
at JLab, HERMES, COMPASS and eRHIC. We have used model I(a) for this comparison. 

So far we calculated the asymmetry using a Gaussian ansatz for the transverse momentum 
dependence of the WW function given by Eq. (\ref{gauss-g}). We also do a comparative study taking a different 
ansatz, namely, the dipole form with $k_0 = 0.7$~GeV.

We have shown the asymmetry as a function of rapidity or COMPASS energy in Fig. 9 
for both Ansatze and using models I(a) and I(b). 
It is seen that the asymmetry does not depend much on the choice of the 
WW function.

\section{SUMMARY AND CONCLUSION}

We have estimated the magnitude of SSA in electroproduction of  $J/\psi$ using 
Weizsacker-Williams  equivalent photon approximation and models of Sivers function 
proposed earlier to estimate the asymmetry in SIDIS and Drell-Yan processes. 
We used the color evaporation model for charmonium production for the numerical estimate of 
the asymmetry. Corresponding to each of the two models of the Sivers function, results 
have been given for two parameterizations of the gluon Sivers function. We have found sizable
asymmetry in the energies of COMPASS, HERMES, JLab and eRHIC. We have used two different ansatz 
for the transverse momentum dependence of the WW function and found that the asymmetry
does not depend much on the choice. 
Our results based on CEM indicate that it may be worthwhile to look at SSA's in 
charmonium production both from the point of view of comparing different models 
of charmonium production as well as comparing the different models of gluon 
Sivers function used for estimating SSA in other processes. 
\section{Acknowledgement}
We thank D. Boer and P. Mulders for helpful comments and discussions.
R.M.G. wishes to acknowledge support from the Department of Science and
Technology, India under Grant No. SR/S2/JCB-64/2007. A. Misra and V.S.R. would like to 
thank Department of Science and Technology, India for financial support under the Grant No. 
SR/S2/HEP-17/2006 and the Department of Atomic Energy-BRNS, India under the Grant No. 2010/37P/47/BRNS.
\vspace{1 cm}


\newpage
 \begin{figure}
\includegraphics[width=0.49\linewidth,angle=0] {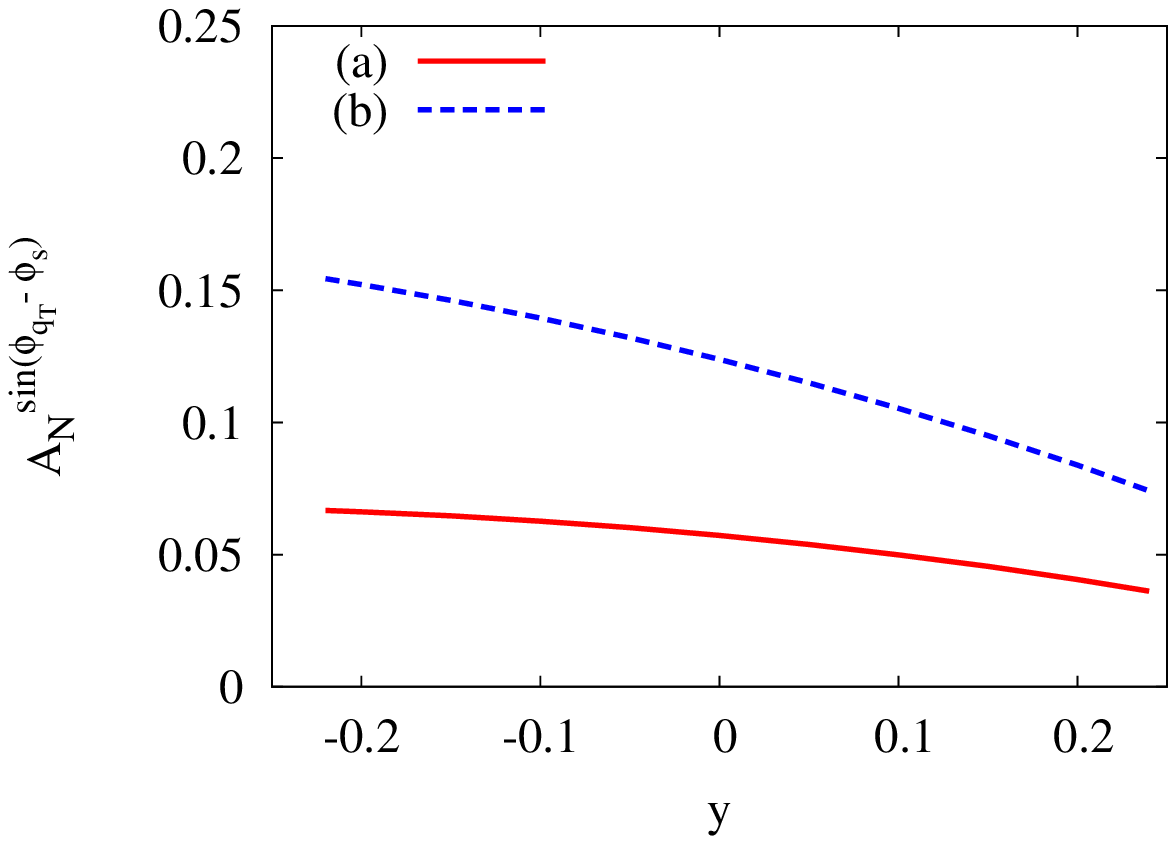}\hspace*{0.2cm}
\includegraphics[width=0.49\linewidth,angle=0] {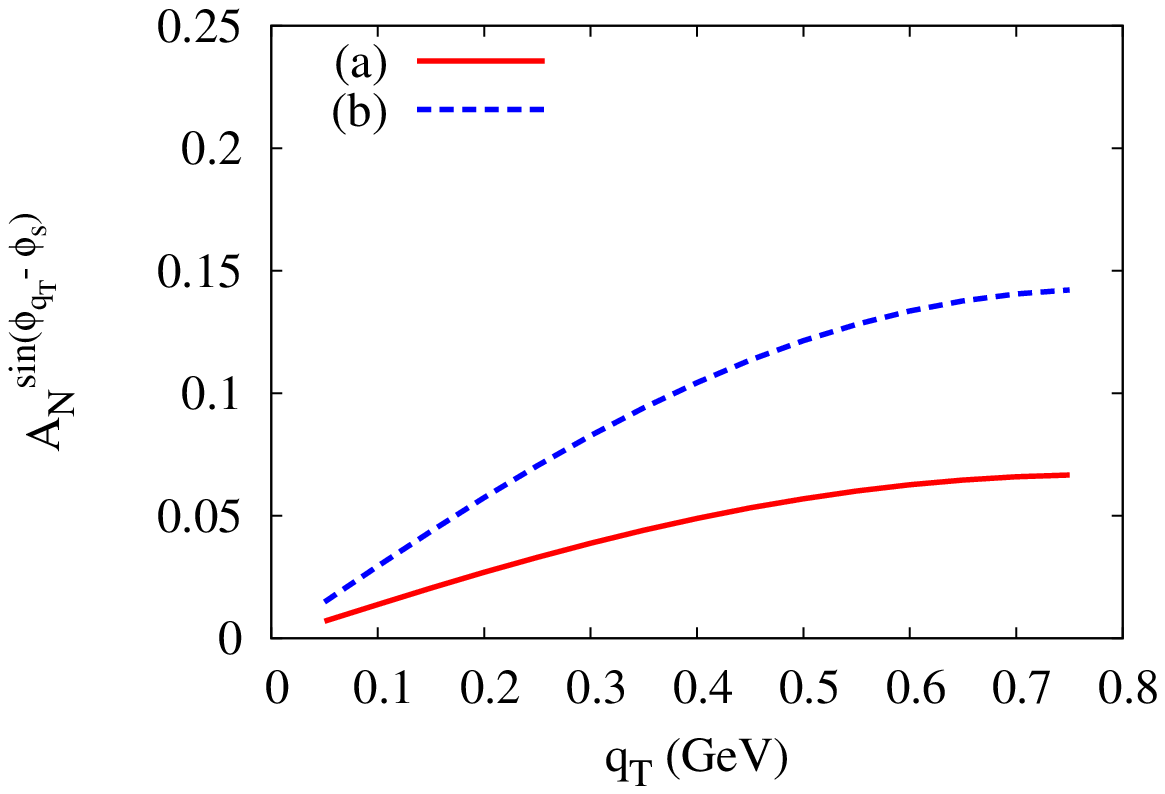}
\caption{(Color online) The single spin asymmetry $A_N^{\sin({\phi}_{q_T}-\phi_S)}$
 for the $e + p^\uparrow \rightarrow e+J/\psi+X$ at JLab as a function of y (left panel) and $q_T$ (right panel). 
The plots are for model I with two parameterizations (a) [solid red line] and (b) [dashed blue line]. 
The integration ranges are $(0 \leq q_T \leq 1)$~GeV and $(0 \leq y \leq 0.25)$. 
 The results are given at $\sqrt s = 4.7$~GeV.}  
\end{figure}
%
 \begin{figure}
\includegraphics[width=0.49\linewidth,angle=0]{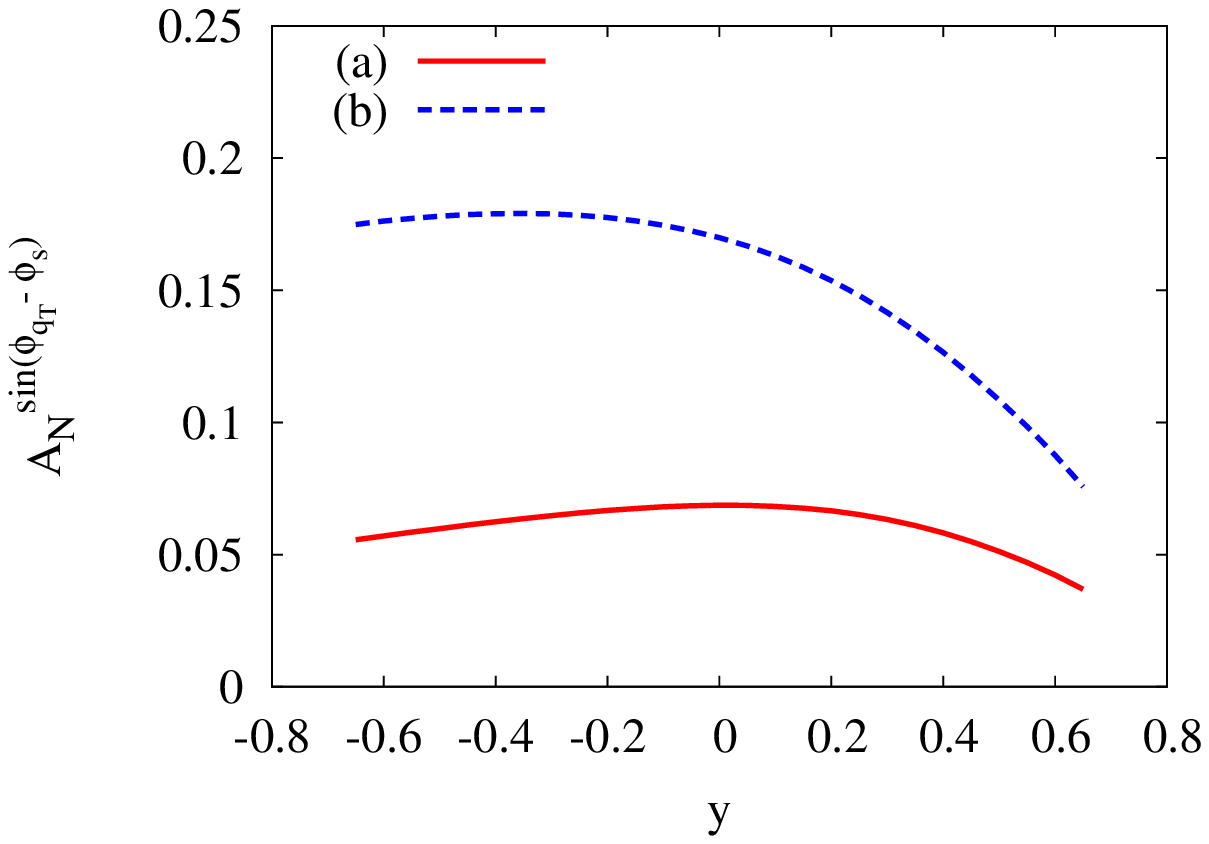}\hspace*{0.2cm}
\includegraphics[width=0.49\linewidth,angle=0]{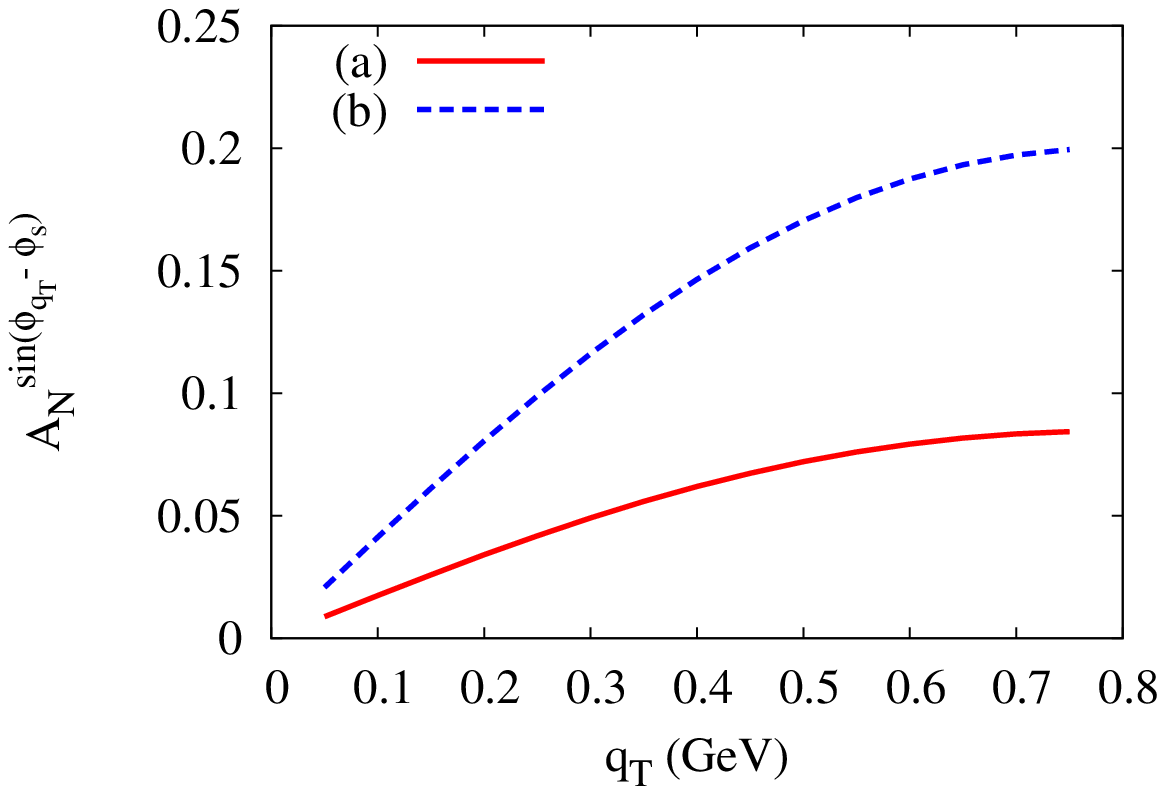}
 \caption{(Color online) The single spin asymmetry $A_N^{\sin({\phi}_{q_T}-\phi_S)}$
 for the $e + p^\uparrow \rightarrow e+J/\psi+X$ at HERMES as a function of y (left panel) and $q_T$ (right panel). 
 The plots are for model I with two parameterizations (a) [solid red line] and (b) [dashed blue line]. 
The integration ranges are $(0 \leq q_T \leq 1)$~GeV and $(0 \leq y \leq 0.6)$. 
 The results are given at $\sqrt s = 7.2$~GeV.}  
\end{figure}
%
 \begin{figure}
\includegraphics[width=0.49\linewidth,angle=0]{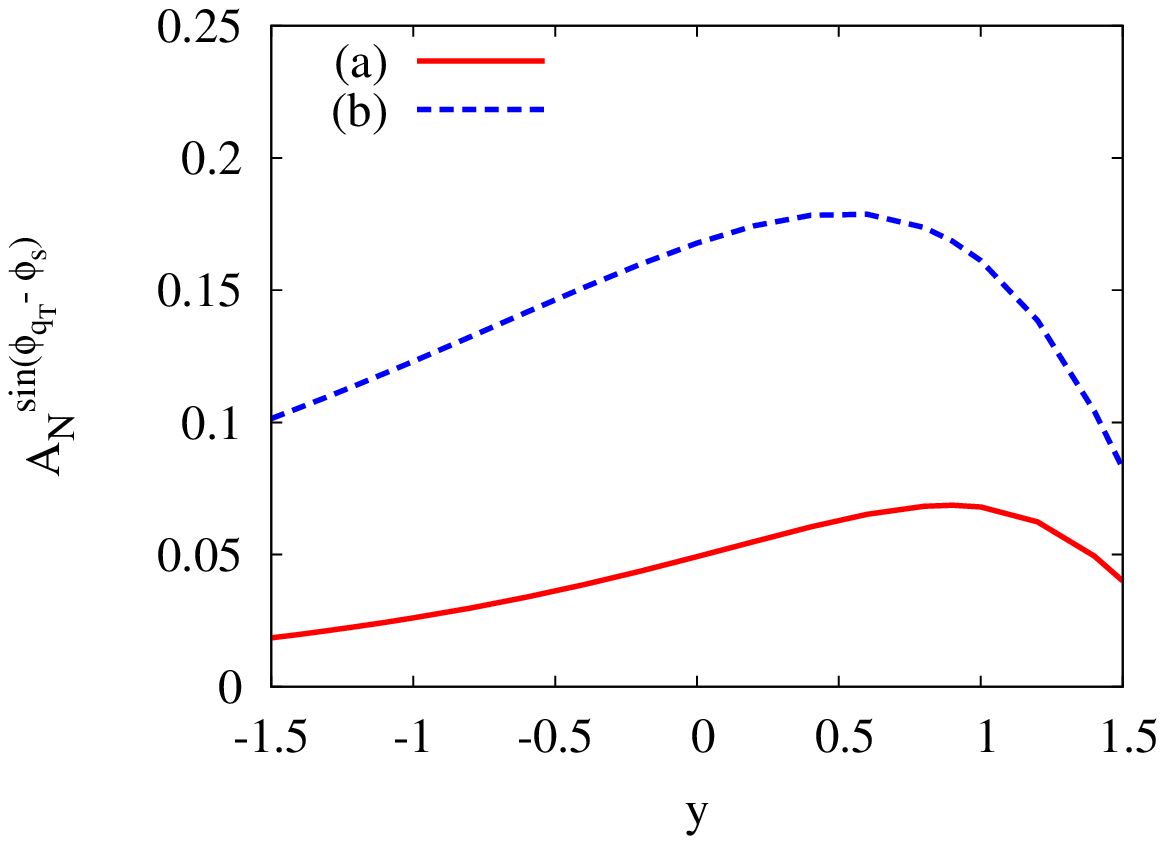}\hspace*{0.2cm}
\includegraphics[width=0.49\linewidth,angle=0]{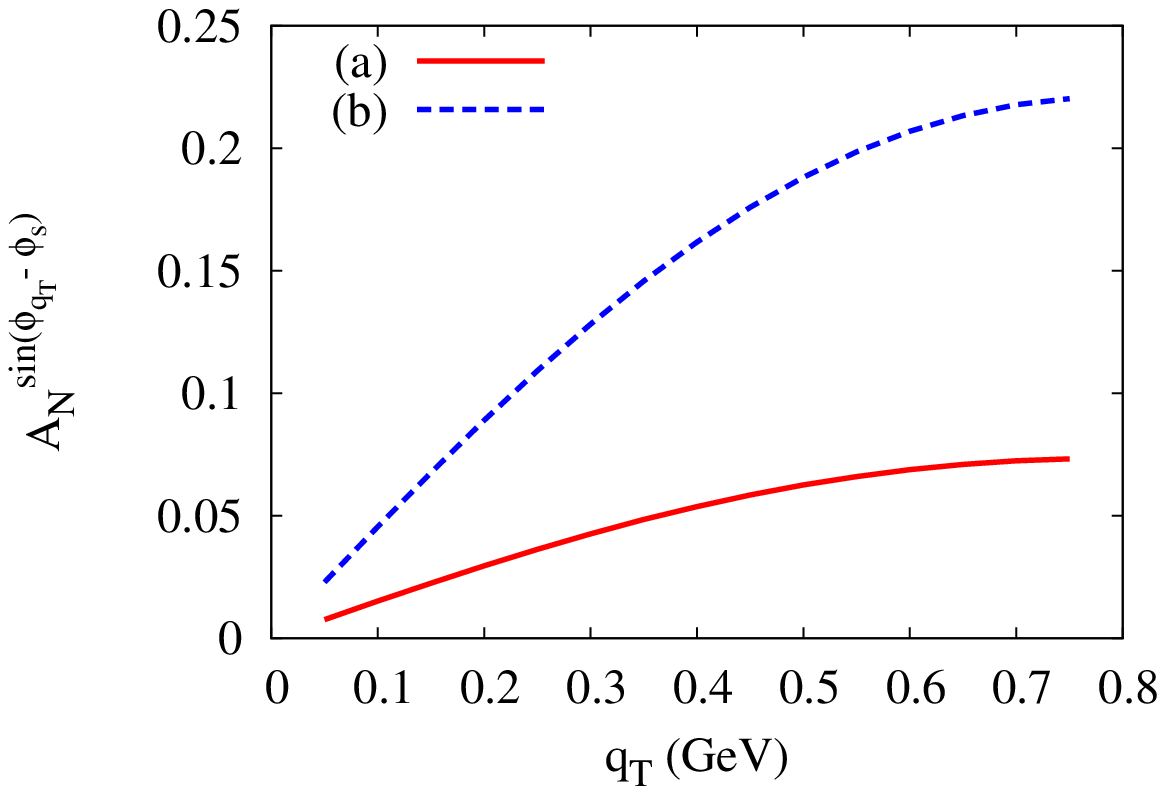}
 \caption{(Color online) The single spin asymmetry $A_N^{\sin({\phi}_{q_T}-\phi_S)}$
 for the $e + p^\uparrow \rightarrow e+J/\psi+X$ at COMPASS  as a function of y (left panel) and $q_T$ (right panel). 
The plots are for model I with two parameterizations (a) [solid red line] and (b) [dashed blue line]. 
The integration ranges are $(0 \leq q_T \leq 1)$~GeV and $(0 \leq y \leq 1)$. 
The results are given at $\sqrt s = 17.33$~GeV.}  
\end{figure}
 \begin{figure}
\includegraphics[width=0.49\linewidth,angle=0]{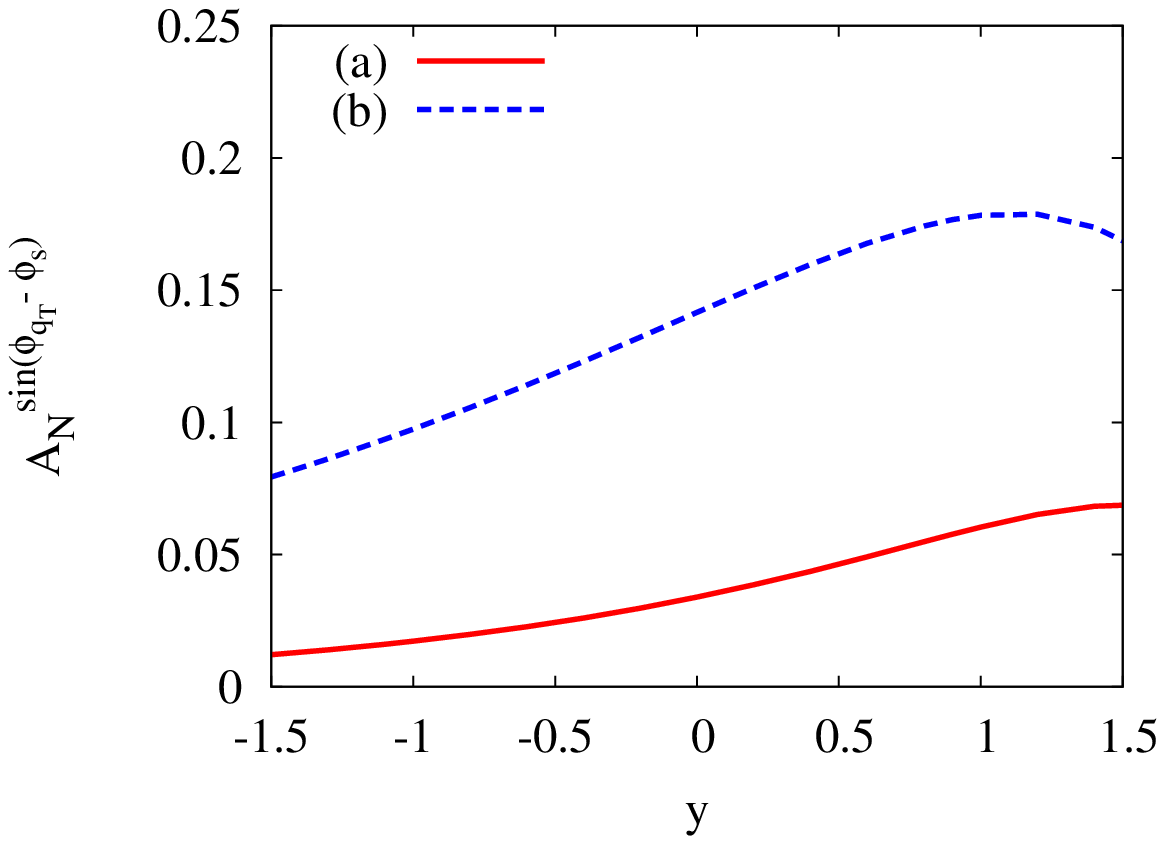}\hspace*{0.2cm}
\includegraphics[width=0.49\linewidth,angle=0]{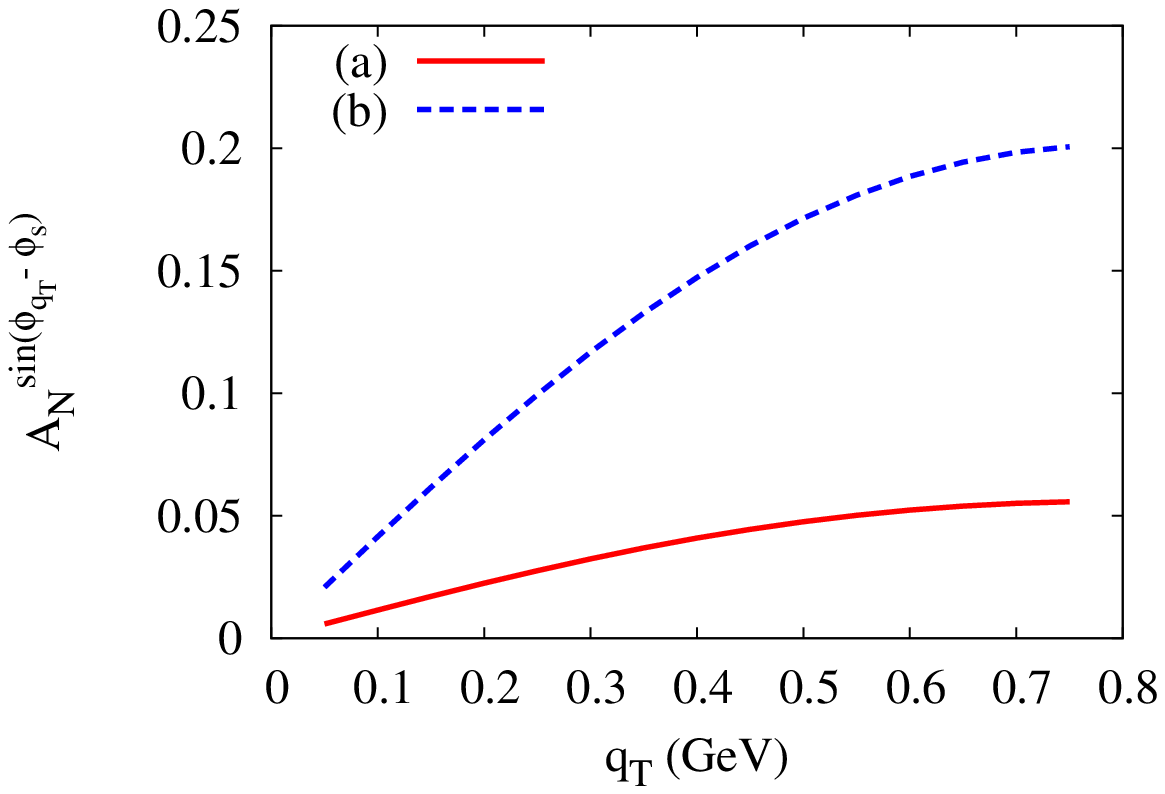}
 \caption{(Color online) The single spin asymmetry $A_N^{\sin({\phi}_{q_T}-\phi_S)}$
 for the $e + p^\uparrow \rightarrow e+J/\psi+X$ at eRHIC as a function of y (left panel) and $q_T$ (right panel). 
The plots are for model I with two parameterizations (a) [solid red line] and (b) [dashed blue line]. 
The integration ranges are $(0 \leq q_T \leq 1)$~GeV and $(0 \leq y \leq 1)$. 
 The results are given at $\sqrt s = 31.6$~GeV.}  
\end{figure}
 \begin{figure}
\includegraphics[width=0.49\linewidth,angle=0]{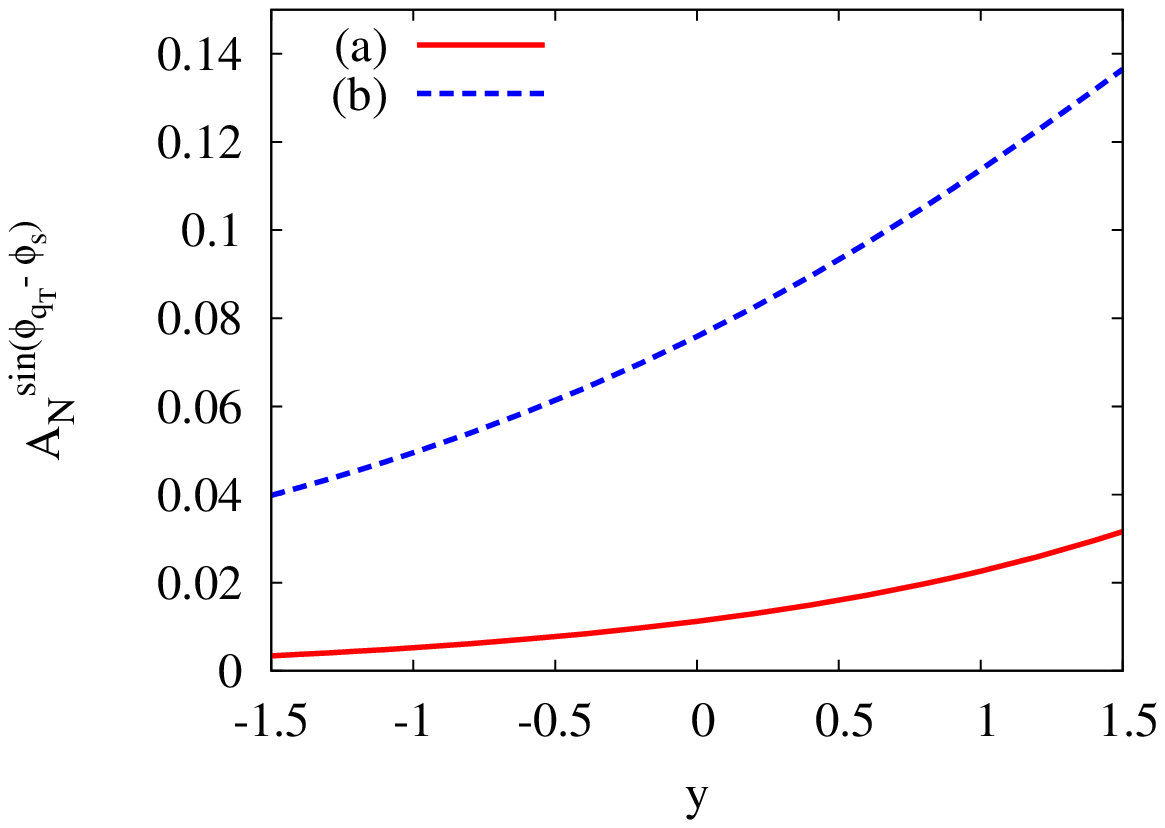}\hspace*{0.2cm}
\includegraphics[width=0.49\linewidth,angle=0]{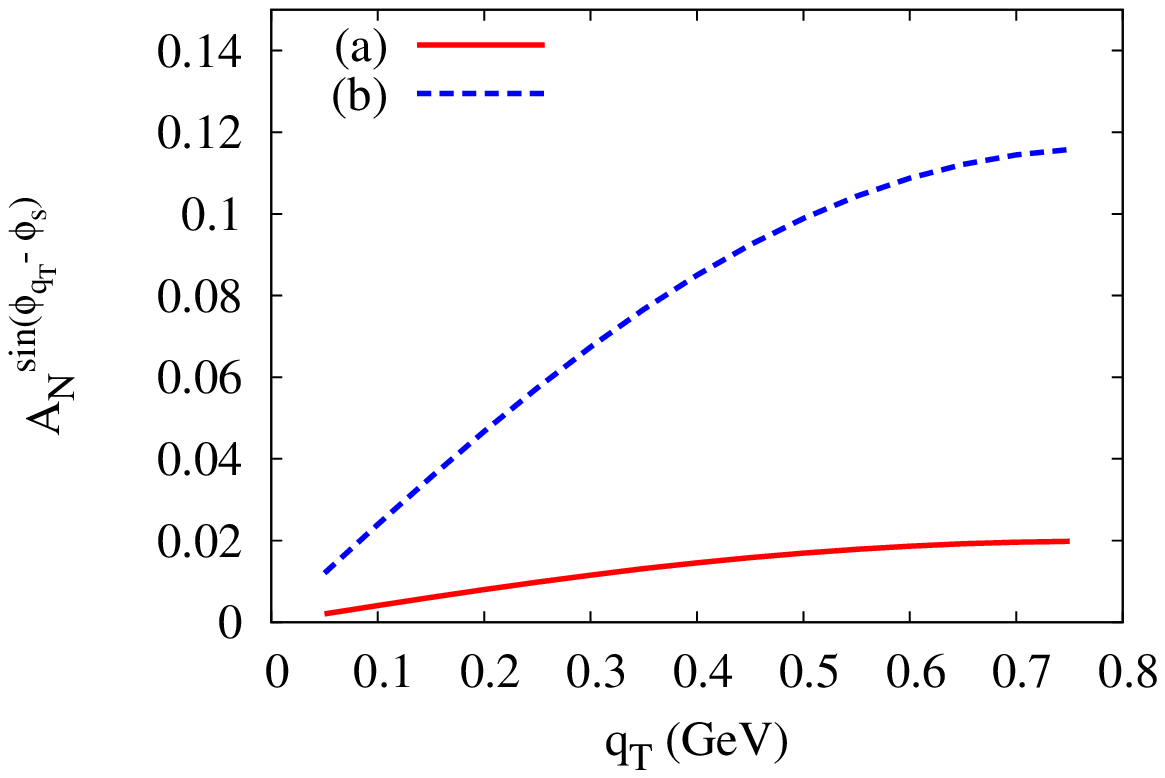}
 \caption{(Color online) The single spin asymmetry $A_N^{\sin({\phi}_{q_T}-\phi_S)}$
 for the $e + p^\uparrow \rightarrow e+J/\psi+X$ at eRHIC as a function of y (left panel) and $q_T$ (right panel). 
The plots are for model I with two parameterizations (a) [solid red line] and (b) [dashed blue line]. 
The integration ranges are $(0 \leq q_T \leq 1)$~GeV and $(0 \leq y \leq 1)$. 
 The results are given at $\sqrt s = 158.1$~GeV.}  
\end{figure}

\begin{figure}
\includegraphics[width=0.49\linewidth,angle=0]{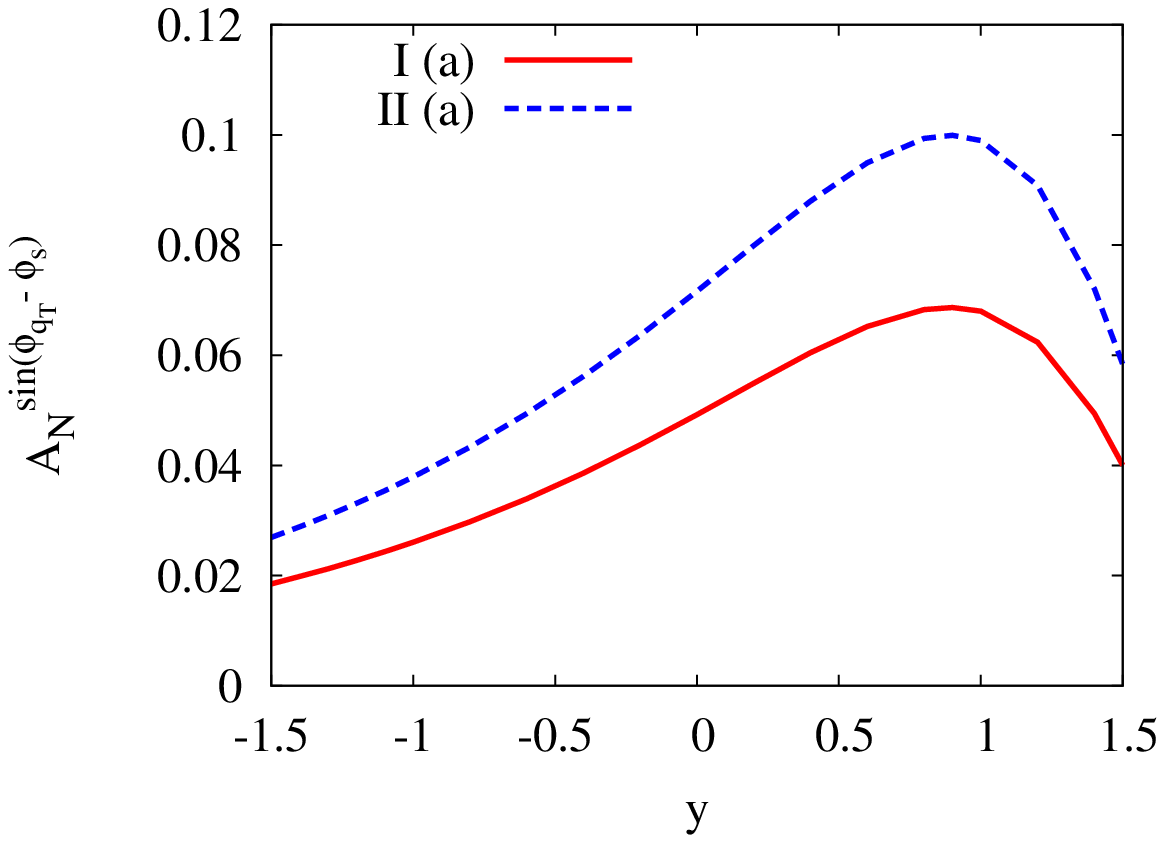}\hspace*{0.2cm} 
\includegraphics[width=0.49\linewidth,angle=0]{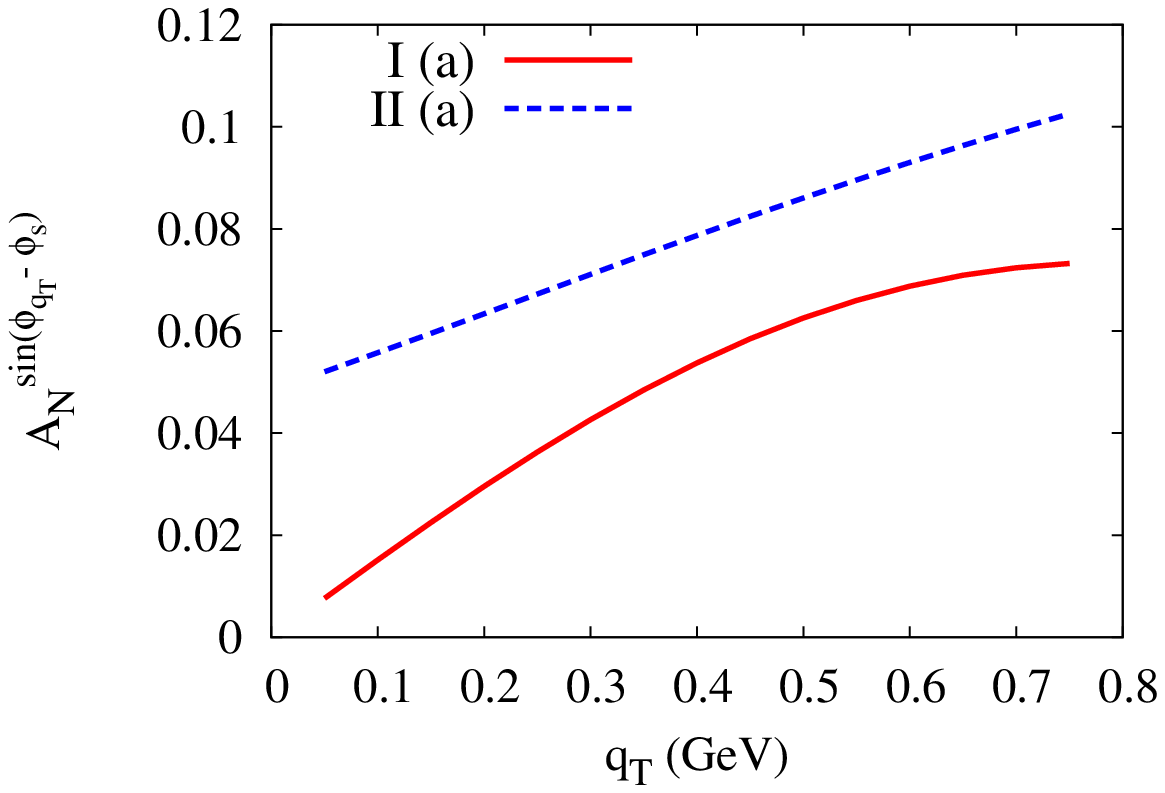}
 \caption{(Color online) The single spin asymmetry $A_N^{\sin({\phi}_{q_T}-\phi_S)}$
for the $e + p^\uparrow \rightarrow e+J/\psi+X$ at COMPASS as a function of y (left panel) and $q_T$ (right panel). 
The plots are for two models I [solid red line] and II [dashed blue line] with parameterization (a). 
The integration ranges are $(0 \leq q_T \leq 1)$~GeV and $(0 \leq y \leq 1)$. 
The results are given at $\sqrt s = 17.33$~GeV.}  
\end{figure}

\begin{figure}
\includegraphics[width=0.49\linewidth,angle=0]{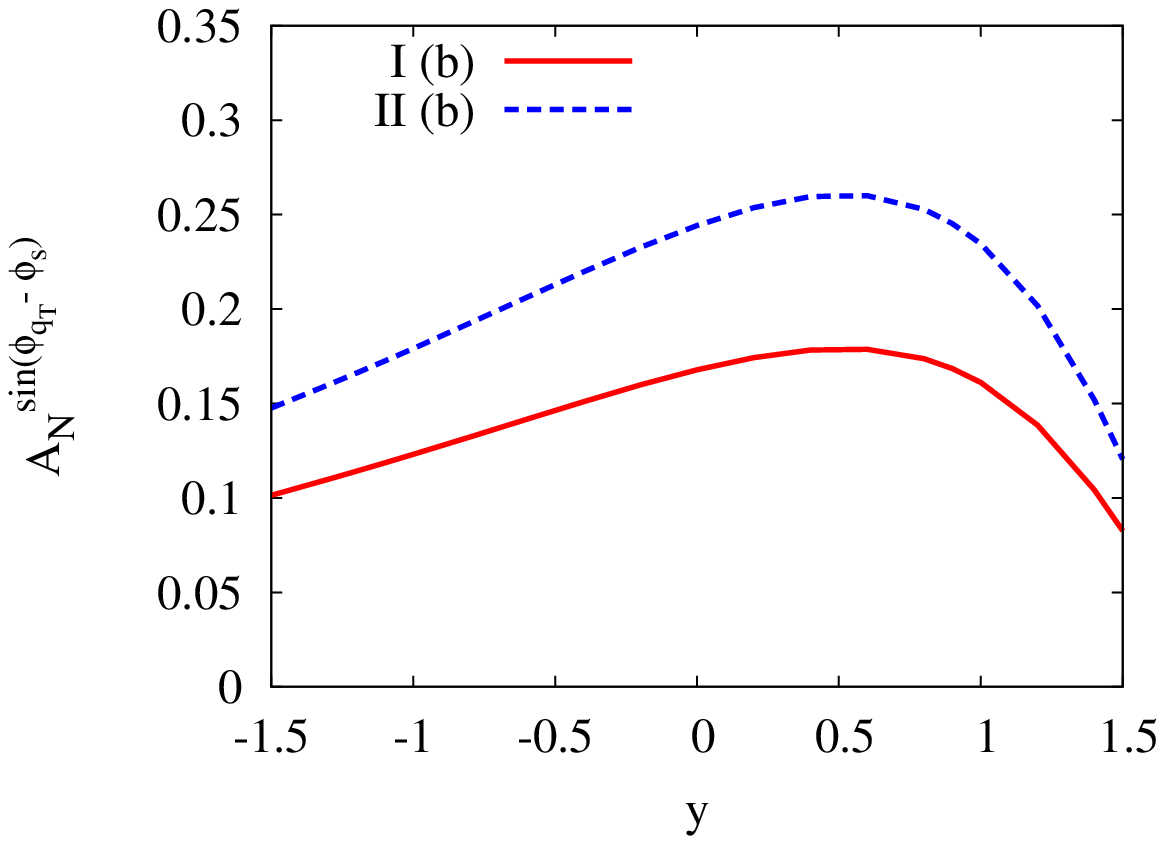}\hspace*{0.2cm} 
\includegraphics[width=0.49\linewidth,angle=0]{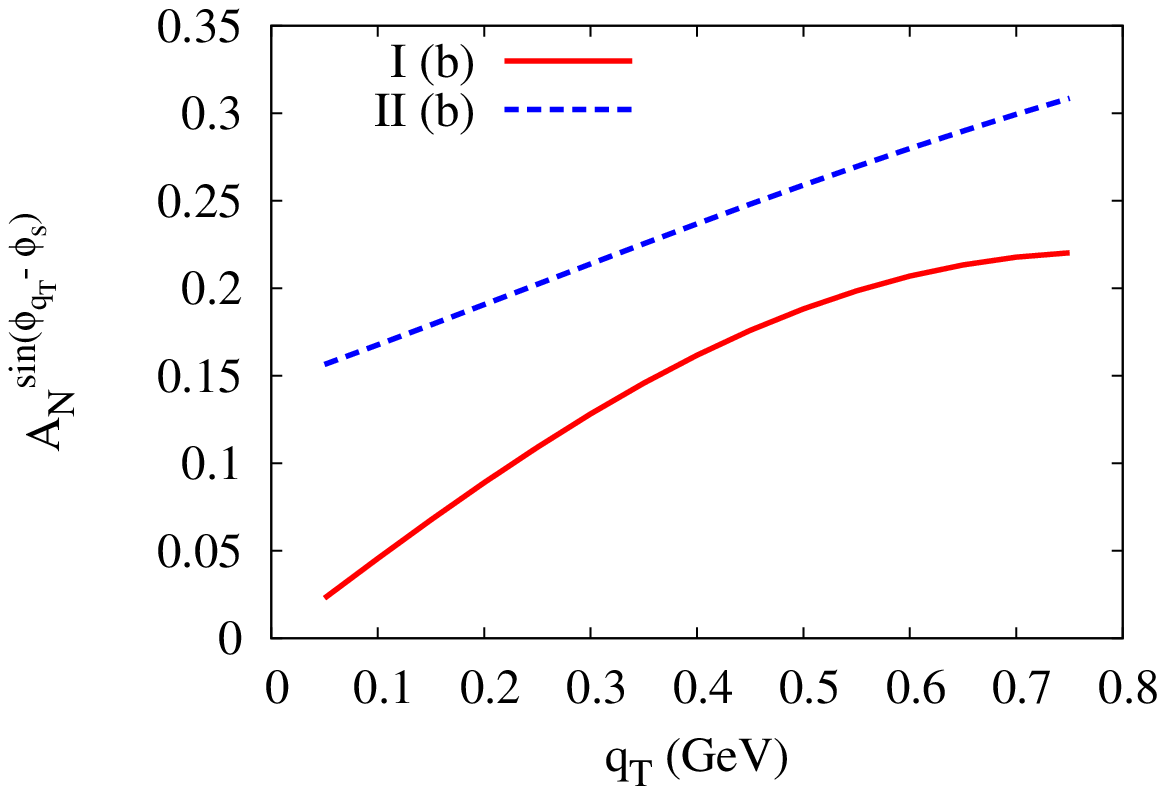}
 \caption{(Color online) The single spin asymmetry $A_N^{\sin({\phi}_{q_T}-\phi_S)}$
for the $e + p^\uparrow \rightarrow e+J/\psi+X$ at COMPASS as a function of y (left panel) and $q_T$ (right panel). 
The plots are for two models  I [solid red line] and II [dashed blue line] with parameterization (b). 
The integration ranges are $(0 \leq q_T \leq 1)$~GeV and $(0 \leq y \leq 1)$. 
The results are given at $\sqrt s = 17.33$~GeV.}  
\end{figure}

\begin{figure}
\includegraphics[width=0.49\linewidth,angle=0]{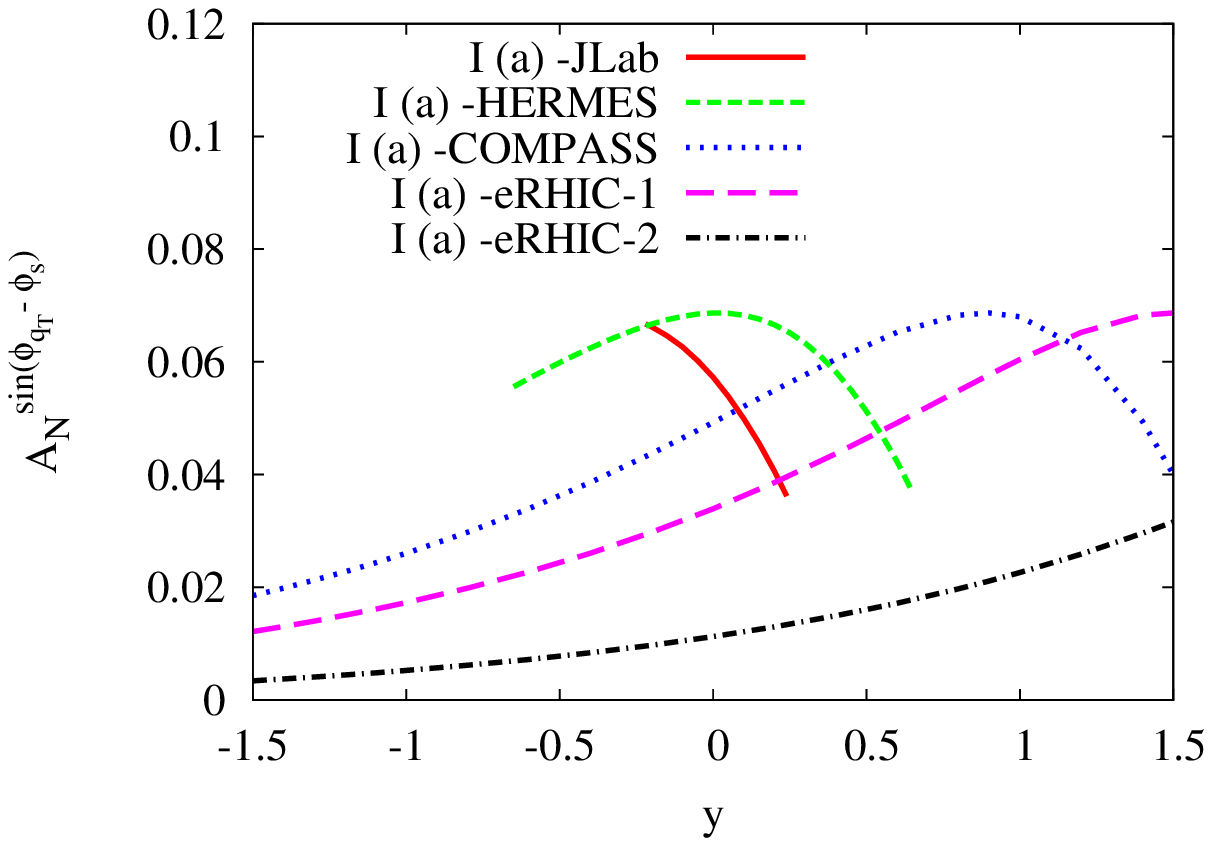}\hspace*{0.2cm} 
\includegraphics[width=0.49\linewidth,angle=0]{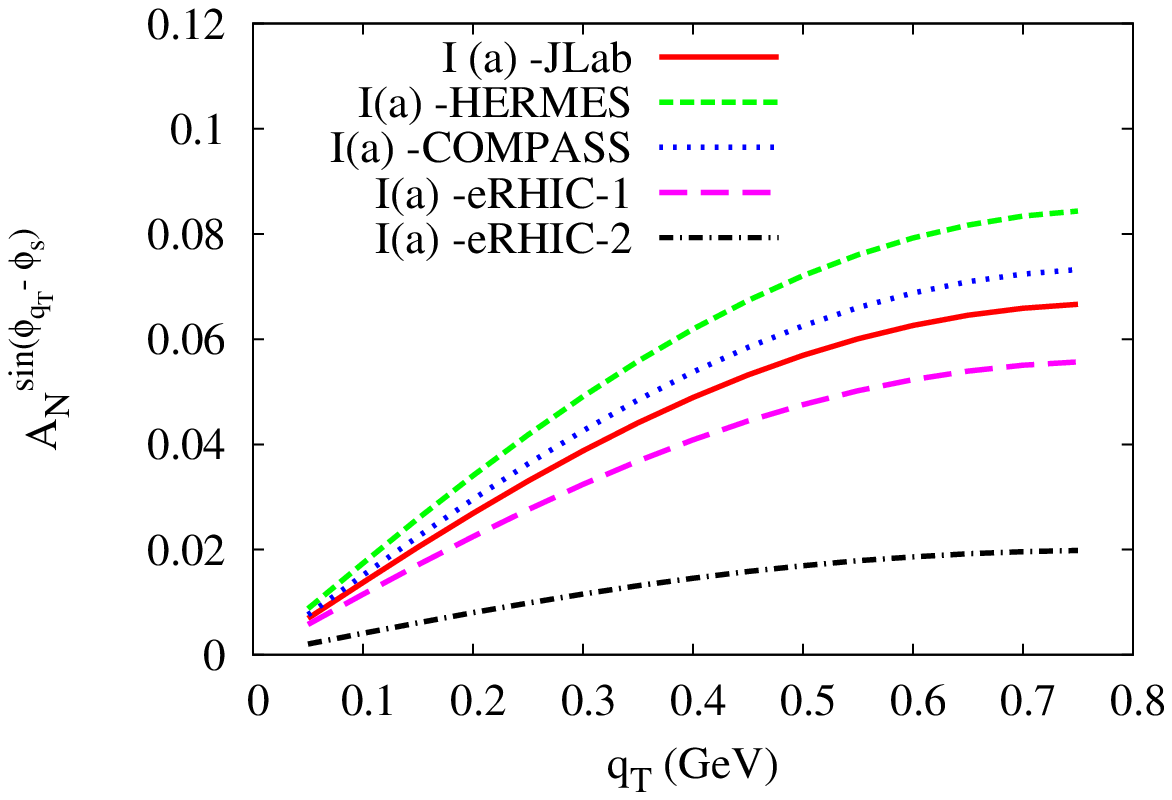}
 \caption{(Color online) The single spin asymmetry $A_N^{\sin({\phi}_{q_T}-\phi_S)}$
for the $e + p^\uparrow \rightarrow e+J/\psi+X$ as a function of y (left panel)  and $q_T$ (right panel). 
The plots are for model I with parameterization (a) compared for JLab ($\sqrt s = 4.7$~GeV) [solid red line], 
HERMES ($\sqrt s = 7.2$~GeV) [dashed green line], COMPASS ($\sqrt s = 17.33$~GeV) [dotted blue line], 
eRHIC-1 ($\sqrt{s}=31.6$~GeV) [long dashed pink line] and eRHIC-2 ($\sqrt{s}=158.1$~GeV) [dot-dashed black line].}  
\end{figure}

\begin{figure}
\includegraphics[width=0.5\linewidth,angle=0]{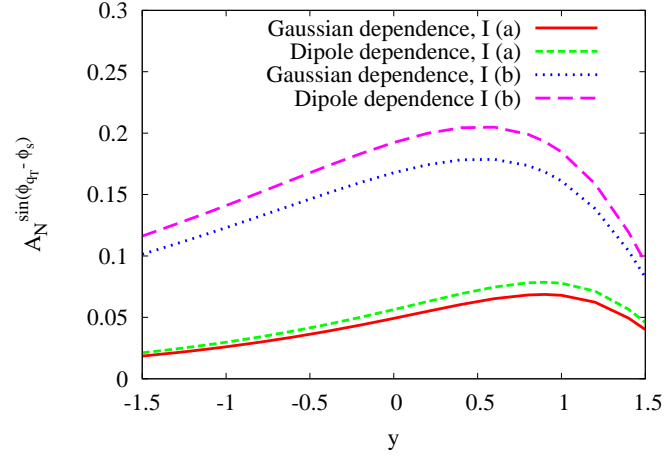} 
\caption{(Color online) The single spin asymmetry $A_N^{\sin({\phi}_{q_T}-\phi_S)}$
for the $e + p^\uparrow \rightarrow e+J/\psi+X$ at COMPASS as a function of y. 
The plots are for model I with two parameterizations (a) [solid red line and dashed green line] and (b) [dotted blue
line and long dashed pink line] compared for Gaussian and dipole $\kt$ dependence of WW function. 
The results are given at $\sqrt s = 17.33$~GeV.}  
\end{figure}

\end{document}